\documentclass[aps,a4paper,amsmath,amssymb]{revtex4}
\usepackage{exscale,latexsym}
\usepackage[dvips]{graphicx}
\begin{document}
\preprint{KLT-\today}
\title{Generalized string theory mapping relations between gravity and gauge theory}
\author{N.~E.~J.~Bjerrum-Bohr}
\email{bjbohr@nbi.dk}
\affiliation{The Niels Bohr Institute,\\
Blegdamsvej 17, DK-2100 Copenhagen, Denmark}
\date{\today}
\begin{abstract}
A previous study of the Kawai, Lewellen and Tye (KLT) relations between gravity and gauge theories, imposed by the relationship of closed and open strings, are here
extended in the light of general relativity and Yang-Mills theory as effective field theories. We discuss the possibility of generalizing the traditional KLT mapping
in this effective setting.
A generalized mapping between the effective Lagrangians of gravity and Yang-Mills theory is presented, and the corresponding operator
relations between gauge and gravity theories at the tree level are further explored. From this generalized mapping remarkable diagrammatic
relations are found, -- linking diagrams in gravity and Yang-Mills theory, -- as well as diagrams in pure effective Yang-Mills theory.
Also the possibility of a gravitational coupling to an antisymmetric field in the gravity scattering amplitude is considered, and
shown to allow for mixed open-closed string solutions, i.e., closed heterotic strings.
\end{abstract}\maketitle
\section{Introduction}
In a former publication~\cite{Bjerrum-Bohr:2003vy} the
implications of the Kawai, Lewellen and Tye (KLT) string
relations~\cite{KLT} were surveyed in the effective field theory
limit of open and closed strings. At such energy scales the tree
scattering amplitudes of strings can be exactly reproduced by the
tree scattering amplitudes originating from an effective
Lagrangian with an infinite number of terms written as a series in
the string tension
$\alpha'$~\cite{Tsey1,Tsey2,Tsey3,Gross:1986iv,Callan:jb,Gross:1985fr,Hull:yi,Sannan:tz}.
Closed strings have a well known natural interpretation as
fundamental theories of gravity -- open string theories turn on
the other hand up as non-abelian Yang-Mills vector field theories.
This gravity/gauge correspondence will be the turning point of our
investigations.

At tree level the KLT string relations are connecting on-shell scattering amplitudes for closed strings with products of
left/right moving amplitudes for open strings. This was first applied by~\cite{Berends} to gain useful information about tree gravity scattering
amplitudes from the much simpler Yang-Mills tree amplitudes. The KLT relations between scattering amplitudes alone give a rather
remarkable non-trivial link between string theories, but in the field theory regime the existence of such relationships is almost astonishing,
and nonetheless valid.

Yang-Mills theory and general relativity being both non-abelian gauge theories have some resemblance, but their dynamical and limiting behaviors at high energy
scales are in fact quite dissimilar. One of the key results of our previous efforts was to show that once the KLT relations were imposed, they uniquely relate the tree limits of
the generic effective Lagrangians of the two theories. The generic effective Lagrangian for gravity will have a restricting tree level connection to that of Yang-Mills theory,
through the KLT relations. In the language of field theories the KLT relations present us with a link between the on-shell field operators in a gravitational theory
as a product of Yang-Mills field theory operators.

Even though the KLT relations only hold explicitly at tree level the factorization of gravity amplitudes into Yang-Mills amplitudes
can be applied with great success in loop calculations. This has been carried out in~\cite{Bern1,Bern2,Bern2b} using
loop diagram cuts and properties such as unitarity of the S-matrix. For a review of such calculations in QCD see~\cite{Bern11}, and also~\cite{Bern0,Dunbar}.
These investigations have also showed the important result that N=8 SUGRA, is less UV divergent than previously believed. In addition, matter sources can be
introduced in the KLT formalism as done in~\cite{Bern4}. We will not consider such loop extensions here, instead we will be more interested in the actual factorization
of tree amplitudes.

Zero string tension factorizations of gravity vertices into gauge vertices were explored at the Lagrangian level in~\cite{Bern3} using
the standard Einstein action: ${\cal L} =\int d^4x \sqrt{-g}R$. Remarkable factorizations of gravity vertices have also been presented in~\cite{Siegel}
using a certain vierbein formalism. The idea that gravity could be factorizable into a product of
independent Yang-Mills theories is indeed very beautiful and in a sense what is promised by the KLT relations.
The product of two independent Yang-Mills vectors ($A^\mu_L$ and $A^\nu_R$) (without internal contractions and interchanges of vector indices) should then be related to the
gravitational tensor field ($g_{\mu\nu}$) through KLT:
\begin{equation}
A_L^{\mu}\times A_R^{\nu} \sim g^{\mu\nu}
\end{equation}
however such a direct relationship, is only realized at the tree {\em amplitude} level. At the Lagrangian level gauge issues mix in
and complicates the matter. Whether this can be resolved or not -- and how -- remains to be understood.
Our goal here will be to gain more insight into the basic tree level factorizations for the amplitudes
generated from the effective action of general relativity and Yang-Mills at order $O(\alpha'^3)$.

Despite common belief, string theory is not the only possible
starting point for the safe arrival at a successful quantum theory
for gravity below the Planck scale. The standard Einstein action
has for decades been known to be non-renormalizable because of
loop divergencies which were impossible to absorb in a
renormalized standard Einstein action~\cite{Veltman,NON,matter}.
Traditionally it has been believed that a field theory description
of general relativity is not possible and that one should avoid
such theories. However general relativity can consistently be
treated as the low energy limit of an effective field theory, and
such an approach leaves no renormalization issues~\cite{Weinberg}.
Furthermore the experimental expectations of Einstein gravity in
the classical limit are still achieved, and corrections from
higher derivative terms in the effective action are negligible in
the classical equations of motion at normal
energies~\cite{Stelle,Simon:ic}. Various field theory calculations
have already successfully been carried out in this
framework~\cite{Donoghue:dn,B1,B2} as well as in~\cite{DB,QED}
where the mixed theory of general relativity and QED is
considered. These calculations have clearly demonstrated that it
is possible to successfully mathematically describe quantum
gravity and gravitational interactions of matter over an enormous
range of energy scales.

An effective field theory thus appears to be the obvious scene for quantum gravity at normal energies -- i.e., below $10^{19}$ GeV. At the Planck energy
the effective action eventually breaks down, leaving an unknown theory at very high
energies; possibly a string theory. From the effective field theory point of view there is no need, however, to assume that the high energy
theory {\it a priori} should be a string theory. The effective field theory viewpoint allows for a broader range of possibilities
than that imposed by a specific field theory limit of a string theory.

The effective field approach for the Yang-Mills theory in four dimensions is not really necessary, because it is already a renormalizable theory -- but no
principles forbid us the possibility of including additional terms in the non-abelian action and this is anyway needed for a d-dimensional $(d > 4)$ Yang-Mills
Lagrangian. By modern field theory principles the non-abelian Yang-Mills action can always be regarded as an effective field theory and treated as such in calculations.

Our aim here will be to gain additional insight in the mapping
process and we will investigate the options of extending the KLT
relations between scattering amplitudes at the effective field
theory level. Profound relations between effective gravity and
Yang-Mills diagrams and between pure effective Yang-Mills diagrams
will also be presented. To which level string theory actually is
needed in the mapping process will be another aspect of our
investigations.

The KLT relations also work for mixed closed string amplitudes and open strings. In order to explore such possibilities we will augment an antisymmetric term,
which is needed for mixed string modes, to the gravitational effective action~\cite{Gross:1985fr}.

The paper is organized as follows. First we review our previous results for the scattering amplitudes; we briefly discuss the theoretical background for the
KLT relations in string theory and make a generalization of the mapping. The mapping solution is then presented, and we give some beautiful diagrammatic
relationships generated by the mapping solution. These relations not only hold between gravity and Yang-Mills theory, but also in pure Yang-Mills theory itself.
We end the paper with a discussion about the implications of such mapping relations between gravity and gauge theory and try to look ahead.
The same conventions as in the previous paper~\cite{Bjerrum-Bohr:2003vy} will be used, i.e., metric $(+ - - -)$ and units $c=\hbar=1$.

\section{The effective actions for general relativity and gauge theory}
In this section we will review the main results of refs.~\cite{Bjerrum-Bohr:2003vy,Tsey1,Tsey2,Tsey3}.
In order to generate a scattering amplitude we need expressions for the generic Lagrangians in gravity and Yang-Mills theory. In principle any
gauge or reparametrization invariant terms could be included -- but it turns out that some terms will be ambiguous under generic
field redefinitions such as:
\begin{equation}\begin{split}
\delta g_{\mu\nu} &= a_1R_{\mu\nu} + a_2g_{\mu\nu}R \ldots\\
\delta A_\mu &= \tilde a_1{\cal D}_\nu F_{\nu\mu} + \ldots
\end{split}\end{equation}

As the S-matrix is manifestly invariant under a field redefinition such ambiguous term terms need not be included in
the generic Lagrangian which give rise to the scattering amplitude, also~\cite{Gross:1986iv,Deser}. The Lagrangian obtained from the reparametrization
invariant terms alone will be sufficient because we are solely considering scattering amplitudes. Including only field
redefinition invariant contributions in the action one can write for the on-shell action of the gravitational fields
(to order $O(\alpha'^3)$\footnote{\footnotesize{For completeness we note; in e.g. six-dimensions there exist an integral relation linking the
term $R^{\mu\nu}_{\ \  \alpha\beta}
R^{\alpha\beta}_{\ \  \lambda\rho} R^{\lambda\rho}_{\ \
\mu\nu}$ with the term $R^{\mu\nu\alpha}_{\ \ \ \beta} R_{\ \
\nu\lambda}^{\beta\gamma} R^{\lambda}_{\
\mu\gamma\alpha}$, in four-dimensions there exist an algebraic relation also linking these terms, see~\cite{Nieuwenhuizen}.
For simplicity we have included all the reparametrization invariant terms in our approach and not looked into the possibility
that some of these terms actually might be further related through an algebraic or integral relation.}}:
\begin{equation}\begin{split}\label{eq9}
{\cal L} &= \frac{2\sqrt{-g}}{\kappa^2}\Big[R + \alpha'\big(a_1
R_{\lambda\mu\nu\rho}^2\big) +(\alpha')^2\big(b_1 R^{\mu\nu}_{\ \
\alpha\beta} R^{\alpha\beta}_{\ \  \lambda\rho} R^{\lambda\rho}_{\
\  \mu\nu} +b_2 (R^{\mu\nu}_{\ \  \alpha\beta}
R^{\alpha\beta}_{\ \  \lambda\rho} R^{\lambda\rho}_{\ \
\mu\nu}-2R^{\mu\nu\alpha}_{\ \ \ \beta} R_{\ \
\nu\lambda}^{\beta\gamma} R^{\lambda}_{\
\mu\gamma\alpha})\big)+\ldots\Big]
\end{split}\end{equation}
while the on-shell effective action in the Yang-Mills case (to order $O(\alpha'^3)$ has the form
\begin{equation}\begin{split}\label{eq7}
{\cal L}_{YM}^L &= -\frac18{\rm tr}\Big[F^L_{\mu\nu}F^L_{\mu\nu} +
\alpha'\big(a^L_1
F^L_{\mu\lambda}F^L_{\lambda\nu}F^L_{\nu\mu}\big)
+(\alpha')^2\big(a^L_3F^L_{\mu\lambda}F^L_{\nu\lambda}F^L_{\mu\rho}
F^L_{\nu\rho}+a^L_4F^L_{\mu\lambda}F^L_{\nu\lambda}F^L_{\nu\rho}F^L_{\mu\rho}
\\&+a^L_5F^L_{\mu\nu}F^L_{\mu\nu}F^L_{\lambda\rho}F^L_{\lambda\rho}
+a^L_6F^L_{\mu\nu}F^L_{\lambda\rho}F^L_{\mu\nu}F^L_{\lambda\rho}\big)
+\ldots \Big]\end{split}\end{equation}
The $L$ in the above equation states that this is the effective Lagrangian for a 'left' moving vector field. To the 'left' field action is associated an
independent 'right' moving action, where 'left' and 'right' coefficients are treated as generally dissimilar. The 'left' and the 'right' theories are
completely disconnected theories and have no interactions. The Yang-Mills coupling constant in the above equation is left out for simplicity. Four additional double trace
terms at order $O(\alpha'^2)$ have been neglected~\cite{Dunbar2}. These terms have a different Chan-Paton structure than the single trace terms.
We need to augment the gravitational effective Lagrangian in order to allow for the heterotic string~\cite{Gross:1985fr}.
As shown in~\cite{Tsey2}, one should in that case add an antisymmetric tensor field coupling:
\begin{equation}
{\cal L}_{\rm Anti} =
\frac{2\sqrt{-g}}{\kappa^2}\Big[-\frac34(\partial_{[\mu}B_{\nu\rho]}+\alpha'c
\omega^{ab}_{[\mu}R^{ab}_{\nu\rho]}+\ldots)(\partial_{[\mu}B_{\nu\rho]}+\alpha'
c \omega^{cd}_{[\mu}R^{cd}_{\nu\rho]}+\ldots) \Big]\end{equation}
where $\omega_{\mu}^{ab}$ represents the spin connection.

This term will contribute to the 4-graviton amplitude.
We will observe how such a term affects the
mapping solution and forces the 'left' and 'right' coefficients
for the Yang-Mills Lagrangians to be different.~\footnote{{See e.g.,
ref.~\cite{Bern:1987bq}, for a discussion of field redefinitions
and the low energy effective action for Heterotic strings.}}

From the effective Lagrangians above one can expand the field invariants and thereby derive the scattering amplitudes. For the 3-point amplitudes this is very
easy as only on-shell terms contribute and we need only consider direct contact terms in the amplitude.

The following 3-point scattering amplitude is generated in gravity:
\begin{equation}\begin{split}
{M_3}&= \kappa \Big[ \zeta_2^{\mu\sigma}\zeta_3^{\mu\rho}
\big(\zeta_1^{\alpha\beta}k_2^{\alpha}k_2^{\beta}\delta^{\sigma\rho}
+\zeta_1^{\sigma\alpha}k_2^{\alpha}k_1^{\rho}
+\zeta_1^{\sigma\alpha}k_3^{\alpha}k_1^{\rho}\big)
+\zeta_1^{\mu\sigma}\zeta_3^{\mu\rho}
\big(\zeta_2^{\alpha\beta}k_3^{\alpha}k_3^{\beta}\delta^{\sigma\rho}
+\zeta_2^{\sigma\alpha}k_1^{\alpha}k_2^{\rho}
+\zeta_2^{\sigma\alpha}k_3^{\alpha}k_2^{\rho}\big)\\&
+\zeta_1^{\mu\sigma}\zeta_2^{\mu\rho}
\big(\zeta_3^{\alpha\beta}k_1^{\alpha}k_1^{\beta}\delta^{\sigma\rho}
+\zeta_3^{\sigma\alpha}k_1^{\alpha}k_3^{\rho}
+\zeta_3^{\sigma\alpha}k_2^{\alpha}k_3^{\rho}\big)
+\alpha'[4a_1\zeta_2^{\mu\sigma}\zeta_3^{\mu\rho}\zeta_1^{\alpha\beta}
k_2^\alpha k_2^\beta k_3^\sigma k_1^\rho
+4a_1\zeta_1^{\mu\sigma}\zeta_3^{\mu\rho}\zeta_2^{\alpha\beta}
k_3^\alpha k_3^\beta k_2^\sigma k_1^\rho\\&
+4a_1\zeta_1^{\mu\sigma}\zeta_2^{\mu\rho}\zeta_3^{\alpha\beta}
k_1^\alpha k_1^\beta k_2^\sigma k_3^\rho]
+(\alpha')^2[12b_1\zeta_1^{\alpha\beta}
\zeta_2^{\gamma\delta}\zeta_3^{\tau\rho}k_1^\tau k_1^\rho
k_2^\alpha k_2^\beta k_3^\gamma k_3^\delta] \Big]
\end{split}\end{equation}
where $\zeta_i^{\mu\nu}$ and $k_i$, $i=1,..,3$ denote the polarization tensors and momenta for the external graviton legs.

For the gauge theory 3-point amplitude one finds:
\begin{equation}\begin{split}
A_{3L} &= -\Big[(\zeta_3\cdot k_1 \zeta_1 \cdot \zeta_2 + \zeta_2\cdot k_3
\zeta_3 \cdot \zeta_1+\zeta_1\cdot k_2 \zeta_2 \cdot \zeta_3)
+ \frac34\alpha'a_1^L\zeta_1\cdot k_2 \zeta_2\cdot k_3 \zeta_3 \cdot k_1\Big]
\end{split}\end{equation}

For the general 4-point amplitude matters are somewhat more complicated. The scattering amplitude will consist both of direct contact terms as well as 3-point contributions
combined with a propagator -- interaction parts. Furthermore imposing on-shell constrains will still leave many non-vanishing terms in the scattering amplitude.
A general polynomial expression for the 4-point scattering amplitude has the form~\cite{Deser,shimada}
\begin{equation}
A_4 \sim \sum_{0\leq n+m+k\leq 2} b_{nmk}s^n t^m u^k
\end{equation}
where $s$, $t$ and $u$ are normal Mandelstam variables and the factor $b_{nmk}$ consist of scalar contractions of momenta and polarizations for the external lines.
The case $n+m+k=2$ corresponds to a particular simple case, where the coefficients $b_{nmk}$ consists only of momentum factors contracted with momentum factors, and polarization indices
contracted with other polarization indices. In the process of comparing amplitudes, we need only to match coefficients for a sufficient part of the amplitude; i.e., for a specific choice
of factors $b_{nmk}$, -- gauge symmetry will then do the rest and dictate that once adequate parts of the amplitudes match, we have achieved matching of the full amplitude.
We will choose the case where $b_{nmk}$ has this simplest form, and the contribution to the gravity 4-point amplitude can then be written:
\begin{equation}\begin{split}
{M_4} &= \frac12
\frac{\kappa^2}{\alpha'}\Big[\zeta_1\zeta_2\zeta_3\zeta_4(z + a_1
z^2 -3b_1(z^3-4xyz) -3b_2(z^3-\frac72xyz) +(a_1^2+3b_1 +
3b_2)(z^3-3xyz)+\frac14c^2(z^3-xyz))\\& +
\zeta_1\zeta_2\zeta_4\zeta_3(y + a_1 y^2
-3b_1(y^3-4xyz)-3b_2(y^3-\frac72xyz) +(a_1^2+3b_1 +
3b_2)(y^3-3xyz)+\frac14c^2(y^3-xyz))\\& +
\zeta_1\zeta_3\zeta_2\zeta_4(x + a_1 x^2
-3b_1(x^3-4xyz)-3b_2(x^3-\frac72xyz) +(a_1^2+3b_1 +
3b_2)(x^3-3xyz)+\frac14c^2(x^3-xyz))\Big]
\end{split}\end{equation}
where we use the definitions: $\zeta_1\zeta_2\zeta_3\zeta_4 = \zeta_1^{\alpha\beta}
\zeta_2^{\beta\gamma}\zeta_3^{\gamma\delta}\zeta_4^{\delta\alpha}$ as well as $x =
-2\alpha'(k_1\cdot k_2)$, $y = -2\alpha'(k_1\cdot k_4)$ and $z=-2\alpha'(k_1\cdot k_3)$.
In the above expression the antisymmetric term (with generic coefficient $c$) needed in the heterotic string scattering amplitude
has been included~\cite{Tsey2,Gross:1985fr}.

The corresponding 4-point gauge amplitude can be written:
\begin{equation}\begin{split}
A_{4L} &= \bigg[
\Big[\zeta_{1324}+\frac{z}{x}\zeta_{1234}+\frac{z}{y}\zeta_{1423}\Big]
+\Big[-\frac{3a^L_1}{8}z(\zeta_{1324}+\zeta_{1234}+\zeta_{1423})\Big]
+\Big[\frac{9(a^L_1)^2}{128}(x(z-y)\zeta_{1234}\\&+y(z-x)\zeta_{1423})\Big]
-\frac14\Big[(\frac12a^L_3)xy\zeta_{1324}-(\frac14a_3^L+2a_6^L)z^2\zeta_{1324}
+(\frac14a^L_3 +\frac12a^L_4)yz\zeta_{1234} +
(\frac14a^L_3 + a^L_5)zx\zeta_{1234}\\&+(\frac12a_4^L+a_5^L)yx\zeta_{1234}
+(\frac14a^L_3 +a_5^L)yz\zeta_{1423}+(\frac14a_3^L+
\frac12a_4^L)xz\zeta_{1423}+(\frac12a_4^L+a_5^L)xy\zeta_{1423}\Big]\bigg]
\end{split}\end{equation}
where $x$, $y$ and $z$ is defined as above, and e.g.
$\zeta_{1234} = (\zeta_1 \cdot \zeta_2)(\zeta_3 \cdot \zeta_4)$
and etc.

These expressions for the scattering amplitudes are quite general.
They relate the general generic effective Lagrangians in the gravity and Yang-Mills case to their 3- and 4-point scattering amplitudes
respectively in any dimension. The next section will be dedicated to showing how the 3-point amplitudes and 4-point amplitudes in gravity
and Yang-Mills have to be related through the KLT relations.

\section{The open-closed string relations}
The KLT relations between closed and open string have already been discussed in ref.~\cite{Bjerrum-Bohr:2003vy} but let us recapitulate
the essentials here. Following conventional string theory~\cite{GSW} the general $M$-point scattering amplitude for a closed string is related to that
of an open string in the following manner:
\begin{equation}\begin{split}
{A}^M_{\rm closed} \sim \sum_{\Pi,\tilde \Pi}
e^{i\pi\Phi(\Pi,\tilde\Pi)}{A}_M^\text{left open}(\Pi)
{A}_M^\text{right open}(\tilde \Pi)
\end{split}\end{equation}
in this expression $\Pi$ and $\tilde\Pi$ corresponds to particular cyclic orderings of the external lines of the open string. While
the $\Pi$ ordering corresponds to a left-moving open string, the $\tilde \Pi$ ordering corresponds to the right-moving string. The
factor $\Phi(\Pi,\tilde \Pi)$ in the exponential is a phase factor chosen appropriately with the cyclic permutations $\Pi$ and
$\tilde \Pi$. In the cases of 3- and 4-point amplitudes, the following specific KLT-relations can be adapted from the $M$-point amplitude:
\begin{equation}\begin{split}\label{eq22}
{M}_{\rm 3 \ gravity}^{\mu\tilde\mu\nu\tilde\nu\rho\tilde\rho}&(1,2,3) =
\kappa{A}^{\mu\nu\rho}_\text{3 L-gauge}(1,2,3)\times
{A}^{\tilde\mu\tilde\nu\tilde\rho}_\text{3 R-gauge}(1,2,3)\\
{M}_{\rm 4 \
gravity}^{\mu\tilde\mu\nu\tilde\nu\rho\tilde\rho\sigma\tilde\sigma}&(1,2,3,4)
= \frac{\kappa^2}{4\pi\alpha'}\sin(\pi x)\times
{A}^{\mu\nu\rho\sigma}_\text{4
L-gauge}(1,2,3,4)\times{A}^{\tilde\mu\tilde\nu\tilde\rho\tilde\sigma}_\text{4
R-gauge}(1,2,4,3)
\end{split}\end{equation}
where $M$ is a tree amplitude in gravity, and we have the color
ordered amplitude $A$ for the gauge theory.

In string theory the specific mapping relations originate through the comparison of open and closed string amplitudes. However considering effective field theories
there is no need to assume anything {\it a priori} about the tree amplitudes. It makes sense to investigate the mapping of scattering amplitudes
in the broadest possible setting. We will thus try to generalize the above mapping relations. In the case of the 4-amplitude mapping it seems that a possibility is
to replace the specific sine function with a general Taylor series e.g.:
\begin{equation}
\frac{\sin(\pi x)}{\pi} \rightarrow xf(x) = x(1+P_1x+P_2x^2+\ldots)
\end{equation}
If this is feasible and what it means for the mapping, will be explored below.

\section{Generalized mapping relations}
Insisting on the ordinary mapping relation for the 3-point amplitude and replacing in the 4-point mapping relation the sine function with a general polynomial the
following relations between the coefficients in the scattering amplitudes are found to be necessary in order for the generalized KLT relations to hold. From the
3-amplitude at order $\alpha'$ we have:

\begin{equation}\begin{split}
3a_1^L+3a_1^R&=16a_1,
\end{split}\end{equation}
while from the 3-amplitude at order $\alpha'^2$ one gets:
\begin{equation}\begin{split}
3a_1^L\, a_1^R &= 64b_1
\end{split}\end{equation}
From the 4-amplitude at order $\alpha'$ we have:
\begin{equation}\begin{split}
16a_1 = 3a_1^L+3a_1^R, \ \ P_1=0
\end{split}\end{equation}
while the 4-amplitude at order $\alpha'^2$ states:
\begin{equation}\begin{split}
6a_5^L+3a_4^L+\frac{27(a_1^L)^2}{16} &= 0,\ \
6a_5^R+3a_4^R+\frac{27(a_1^R)^2}{16} = 0,
\end{split}\end{equation}
together with the equations
\begin{equation}\begin{split}
24c^2+96a_1^2&= 6a_3^L+3a_3^R+18a_4^L+12a_5^L+24a_6^R+\frac{81(a_1^L)^2}{8}+96P_2+18(a_1^R+a_1^L)P_1,\\
24c^2+96a_1^2&= 3a_3^L+6a_3^R+18a_4^R+12a_5^R+24a_6^L+\frac{81(a_1^R)^2}{8}+96P_2+18(a_1^R+a_1^L)P_1,\\
24c^2+96a_1^2&= 6a_3^L-3a_3^R-6a_4^L-36a_5^L-12a_5^R-\frac{27}{8}(a_1^L)^2+\frac{27}{8}(a_1^R)^2+96P_2-18(a_1^R+a_1^L)P_1,\\
24c^2+96a_1^2&=-3a_3^L+6a_3^R-6a_4^R-12a_5^L-36a_5^R+\frac{27}{8}(a_1^L)^2-\frac{27}{8}(a_1^R)^2+96P_2-18(a_1^R+a_1^L)P_1,
\end{split}\end{equation}
and furthermore
\begin{equation}\begin{split}
96b_1+48b_2+16c^2&=4a_3^L+5a_3^R+2a_4^L+6a_4^R+8a_6^L
+\frac{9(a_1^R)^2}{4}+\frac{9a_1^L\, a_1^R}{2}+96P_2+18(a_1^R+a_1^L)P_1,\\
96b_1+48b_2+16c^2&=4a_3^R+5a_3^L+2a_4^R+6a_4^L+8a_6^R
+\frac{9(a_1^L)^2}{4}+\frac{9a_1^L\,a_1^R}{2}+96P_2+18(a_1^R+a_1^L)P_1,
\end{split}\end{equation}
as well as
\begin{equation}\begin{split}
96a_1^2-96b_1-48b_2+8c^2& = a_3^L+2a_4^R-12a_5^L-12a_5^R-\frac{9a_1^L\, a_1^R}{2}+\frac{9((a_1^L)^2+(a_1^R)^2)}{8}+64P_2-6(a_1^R+a_1^L)P_1,\\
96a_1^2-96b_1-48b_2+8c^2& = 2a_4^R-4a_5^L-12a_5^R+8a_6^L-\frac{9a_1^L\, a_1^R}{2}+\frac{9((a_1^L)^2+(a_1^R)^2)}{8}+32P_2,\\
96a_1^2-96b_1-48b_2+8c^2& = 2a_4^L-12a_5^L-4a_5^R+8a_6^R-\frac{9a_1^L\, a_1^R}{2}+\frac{9((a_1^L)^2+(a_1^R)^2)}{8}+32P_2,\\
96a_1^2-96b_1-48b_2+8c^2& =
a_3^R+2a_4^L-12a_5^L-12a_5^R-\frac{9a_1^L\,a_1^R}{2}+\frac{9((a_1^L)^2+(a_1^R)^2)}{8}+64P_2-6(a_1^R+a_1^L)P_1.
\end{split}\end{equation}
These equations are found by relating similar scattering components, e.g., the product resulting from the generalized KLT relations: $\zeta_{1234}\zeta_{1324}y^2x$ on the gauge side,
with $\zeta_1\zeta_2\zeta_4\zeta_3 y^2x$ on the gravity side.
The relations can be observed to be a generalization of the mapping equations we found in ref.~\cite{Bjerrum-Bohr:2003vy}. Allowing for a more general mapping
and including the antisymmetric term needed in the case of an heterotic string generates additional freedom in the mapping equations. The generalized equations can still
be solved and the solution is unique. One one ends up with the following solution~\footnote{{We have employed an algebraic
equation solver {\em Maple}, to solve the equations, (Maple and Maple V are registered trademarks of Maple Waterloo Inc.)}}:
\begin{eqnarray}
a_1^L=\frac8{3}a_1\pm \frac43c,&&
a_1^R=\frac8{3}a_1\mp \frac43c,\\
a_3^L=-8P_2,&&
a_3^R=-8P_2,\\
a_4^L=-4P_2,&&
a_4^R=-4P_2,\\
a_5^L=\mp 2a_1c-2a_1^2-\frac12c^2+2P_2,&&
a_5^R=\pm 2a_1c-2a_1^2-\frac12c^2+2P_2,\\
a_6^L=\pm 2a_1c+2a_1^2+\frac12c^2+P_2,&&
a_6^R=\mp 2a_1c+2a_1^2+\frac12c^2+P_2,\\
b_1=\frac{1}{3}a_1^2-\frac{1}{12}c^2,&&
b_2=\frac{2}{3}a_1^2-\frac16c^2.
\end{eqnarray}
This is the unique solution to the generalized mapping relations.
As seen, the original KLT solution ref.~\cite{Bjerrum-Bohr:2003vy} is still contained but the generalized mapping solution is not as constraining as the
original KLT solution was. In fact now one can freely choose $c$, and $P_2$ as well as $a_1$. Given a certain gravitational action
with or without the possibility of terms needed for heterotic strings, one can choose between different mappings from the gravitational Lagrangian to the given Yang-Mills
action. Traditional string solutions are contained in this and are possible to reproduce -- but the solution space for the generalized solution is broader and allows
seemingly for a wider range of possible effective actions on the gravity and the Yang-Mills side. It is important to note that this does not imply that the coefficients in the effective
actions can be chosen freely -- the generalized KLT relations still present rather restricting constraints on the effective Lagrangians. To which extend one may be able to reproduce
the full solution space by string theory is not definitely answered. Clearly superstrings cannot reproduce the full solution space because of spacetime supersymmetry
which does not allow for terms in the effective action like ${\rm tr}(F_{\mu\nu}F_{\nu\alpha}F_{\alpha\mu})$ on the open string side and
$R^{\mu\nu}_{\alpha\beta}R^{\alpha\beta}_{\gamma\lambda}R^{\gamma\lambda}_{\mu\nu}$ the gravity side~\cite{grisaru1,grisaru2,grisaru3}. For non-supersymmetric string theories
constrains on the effective actions such as the above does not exist, and it is therefore possible that in this case some parts or all of the solution space {\it might} be reproduced
by the variety of non-supersymmetric string theories presently known. It is e.g. observed that the bosonic non-supersymmetric string solution in fact covers parts of the solution space
not covered by the supersymmetric string solution.

The possibility of heterotic strings on the gravity side, i.e., a
non-vanishing $c$, will as observed always generate dissimilar
'left' and 'right' Yang-Mills coefficients. I.e., for nonzero $c$,
e.g., $a_1^L \neq a_1^R$, $a_5^L \neq a_5^R$ and $a_6^L \neq
a_6^R$. It is also seen that for $c=0$ that 'left' is equivalent
to 'right'. The coefficients $a_3^L$, $a_3^R$ and $a_4^L$, $a_4^R$ are
completely determined by the coefficient $P_2$. In the generalized
mapping relation the only solution for the coefficient $P_1$ is
zero.

The KLT or generalized KLT mapping equations can be seen as coefficient constraints linking the generic terms in the gauge/gravity Lagrangians.
To summarize the gravitational Lagrangian has to take the following form, dictated by the generalized KLT relations:
\begin{equation}\begin{split}
{\cal L} &= \frac{2\sqrt{-g}}{\kappa^2}\Big[R + \alpha'\Big(a_1
R_{\lambda\mu\nu\rho}^2\Big) +(\alpha')^2\Big(\big(\frac{1}{3}a_1^2-\frac{1}{12}c^2\big) R^{\mu\nu}_{\ \
\alpha\beta} R^{\alpha\beta}_{\ \  \lambda\rho} R^{\lambda\rho}_{\
\  \mu\nu} \\&+\big(\frac{2}{3}a_1^2-\frac16c^2\big) \big(R^{\mu\nu}_{\ \  \alpha\beta}
R^{\alpha\beta}_{\ \  \lambda\rho} R^{\lambda\rho}_{\ \
\mu\nu}-2R^{\mu\nu\alpha}_{\ \ \ \beta} R_{\ \
\nu\lambda}^{\beta\gamma} R^{\lambda}_{\
\mu\gamma\alpha}\big)\Big)-\frac34\Big(\partial_{[\mu}B_{\nu\rho]}+\alpha' c
\omega^{ab}_{[\mu}R^{ab}_{\nu\rho]}+\ldots\Big)^2+\ldots\Big]
\end{split}\end{equation}
and the corresponding 'left' or 'right' Yang-Mills action are then forced to be:
\begin{equation}\begin{split}
{\cal L}_{YM} &= -\frac18{\rm tr}\Big[F^L_{\mu\nu}F^L_{\mu\nu} +
\alpha'\Big(\big(\frac8{3}a_1\pm \frac43 c \big)
F^L_{\mu\lambda}F^L_{\lambda\nu}F^L_{\nu\mu}\Big)
+(\alpha')^2\Big(-8P_2F^L_{\mu\lambda}F^L_{\nu\lambda}F^L_{\mu\rho}
F^L_{\nu\rho}-4P_2F^L_{\mu\lambda}F^L_{\nu\lambda}F^L_{\nu\rho}F^L_{\mu\rho}
\\&+\big(\mp 2a_1 c-2a_1^2+2P_2-\frac12c^2\big)F^L_{\mu\nu}F^L_{\mu\nu}F^L_{\lambda\rho}F^L_{\lambda\rho}
+\big(\pm 2a_1 c+2a_1^2+\frac12c^2+P_2\big)F^L_{\mu\nu}F^L_{\lambda\rho}F^L_{\mu\nu}F^L_{\lambda\rho}\Big)
+\ldots \Big]\end{split}\end{equation}
where 'left' and 'right,' reflects opposite choices of signs, in the above equation.
One sees that the Yang-Mills Lagrangian is fixed once a gravitational action is chosen; the only remaining freedom in the
Yang-Mills action is then the choice of $P_2$, corresponding to different mappings.

It is directly seen that the graviton 3-amplitude to order $\alpha'$ is re-expressible in
terms of 3-point Yang-Mills amplitudes. This can be expressed diagrammatically in the following way:

\begin{figure}[h]
\includegraphics{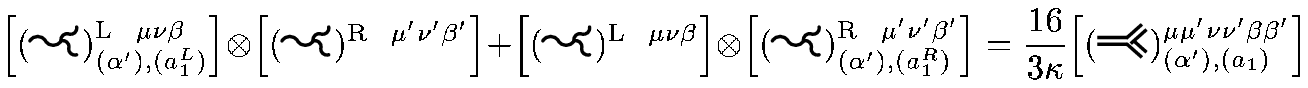}
\caption{An diagrammatical expression for the generalized mapping
of the 3-point gravity amplitude into the product of Yang-Mills
amplitudes at order $\alpha'$.}
\end{figure}

\noindent At order $\alpha'^2$ we have:

\begin{figure}[h]
\includegraphics{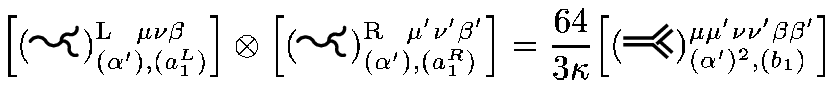}
\caption{An diagrammatical expression for the generalized mapping
of the 3-point gravity amplitude into the product of Yang-Mills
amplitudes at order $\alpha'^2$.}
\end{figure}

At the amplitude level this factorization is no surprise; this is just what the KLT relations tell us. At the
Lagrangian level things usually get more complicated and no factorizations of gravity vertex rules are readily available.
At order $O(\alpha')$ the factorization of gravity vertex rules was investigated in ref.~\cite{Bern3}. An interesting task
would be to continue this analysis to order $O(\alpha'^3)$, and investigate the KLT relations for effective actions directly
at the Lagrangian vertex level. Everything is more complicated in the 4-point case as contact and
non-contact terms mix in the mapping relations. This originates
from the fact that we are actually relating S-matrix
elements. Diagrammatically the 4-point generalized KLT relation at order $\alpha'$ is presented below:
\begin{figure}[h]
\includegraphics{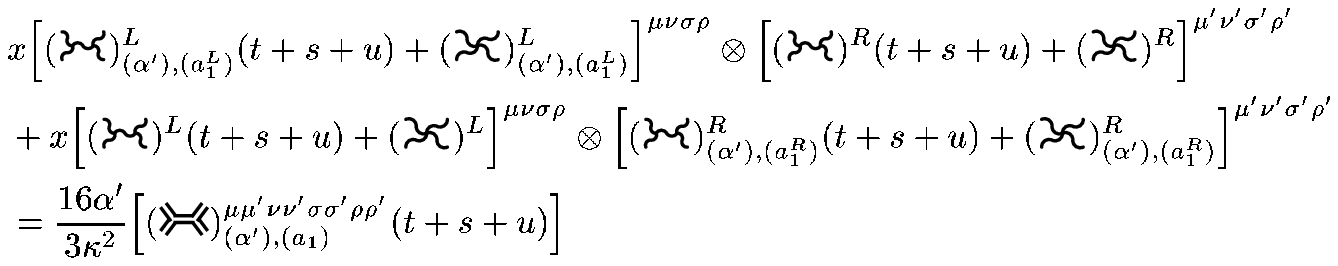}
\caption{The diagrammatical expression for the generalized mapping of the 4-point gravity amplitude as a product of Yang-Mills amplitudes at order $\alpha'$.}
\end{figure}

\noindent This relation is essentially equivalent to the 3-vertex relation at order $\alpha'$. The 4-point
relation at order $\alpha'$ rules out the possibility of a $P_1$ term in the generalized mapping. The full KLT relation at order $\alpha'^2$ is:
\begin{figure}[h]
\includegraphics{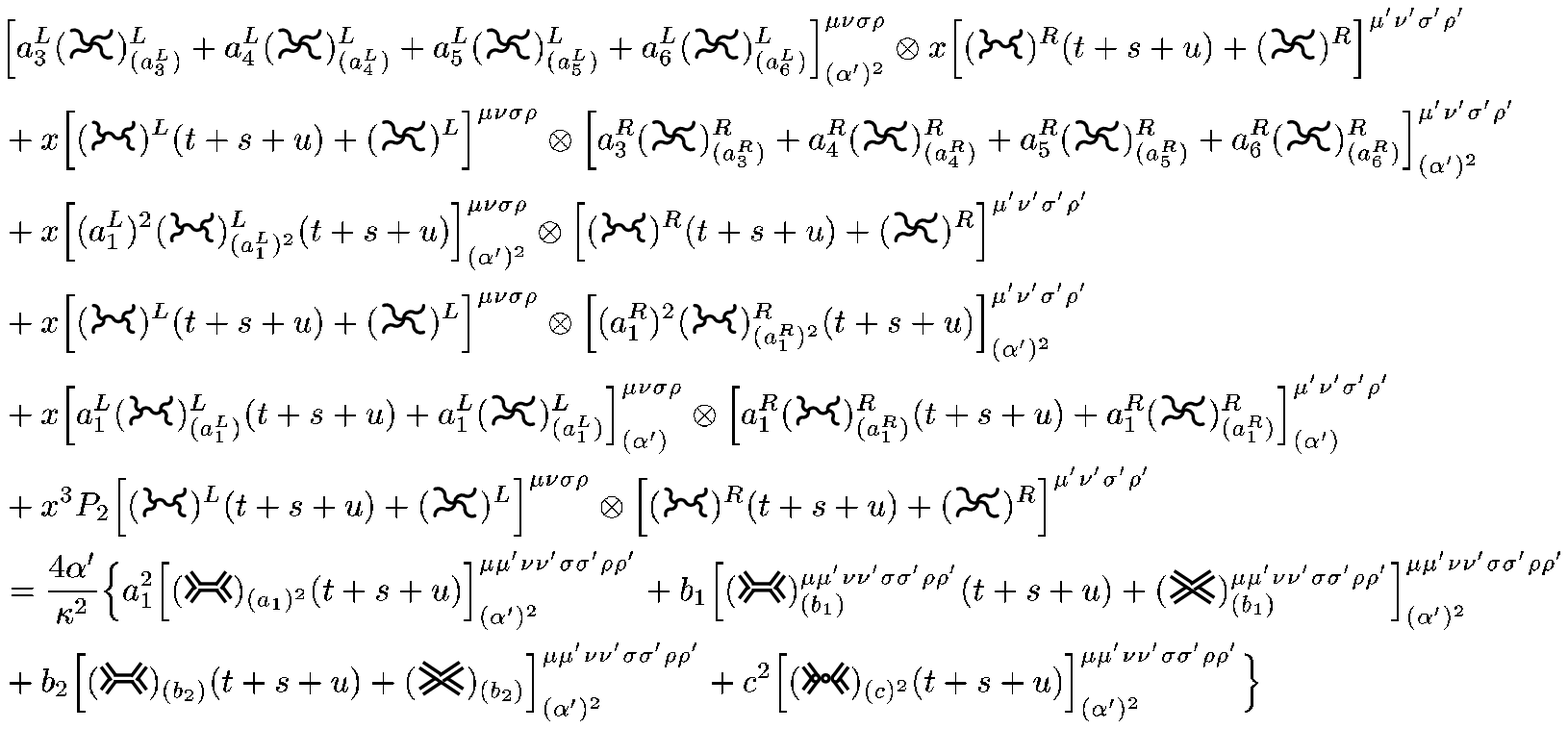}
\caption{The diagrammatical expression for the generalized mapping of the 4-point gravity amplitude into the product of Yang-Mills amplitudes at order $\alpha'^2$.}
\end{figure}

\noindent The coefficients in the above expression need to be taken in agreement with the generalized KLT solution in order for this identity to hold.
A profound consequence of the KLT relations is that they link the sum of a certain class of diagrams in Yang-Mills theory with a corresponding sum of diagrams
in gravity, for very specific values of the constants in the Lagrangian. At glance we do not observe anything manifestly about the decomposition of e.g., vertex rules, -- however
this should be investigated more carefully before any conclusions can be drawn. The coefficient equations cannot directly be transformed into relations between diagrams, -- this can be
seen by calculation.
However reinstating the solution for the coefficients it is possible to turn the above equation for the 4-amplitude at order $\alpha'^2$ into interesting statements
about diagrams. Relating all $P_2$ terms gives e.g., the following remarkable diagrammatic statement:
\begin{figure}[h]
\includegraphics{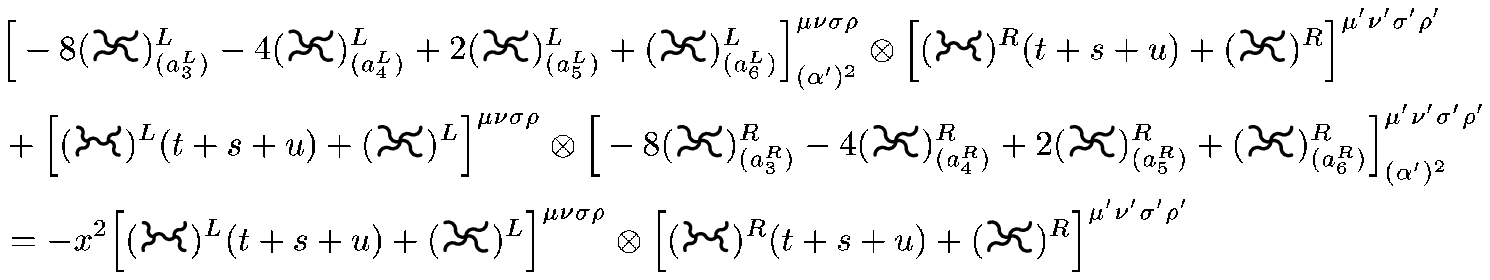}
\caption{A diagrammatic relationship on the Yang-Mill side for the tree operators in the effective action between operators of unit order and order $\alpha'^2$.}
\end{figure}
\newpage
\noindent The surprising fact is that from the relations which apparently link gravity and Yang-Mills theory only, one can eliminate the gravity part to obtain
relations entirely in pure Yang-Mills theory.

\noindent Similarly, relating all contributions with $a_1^2$ gives:

\begin{figure}[h]
\includegraphics{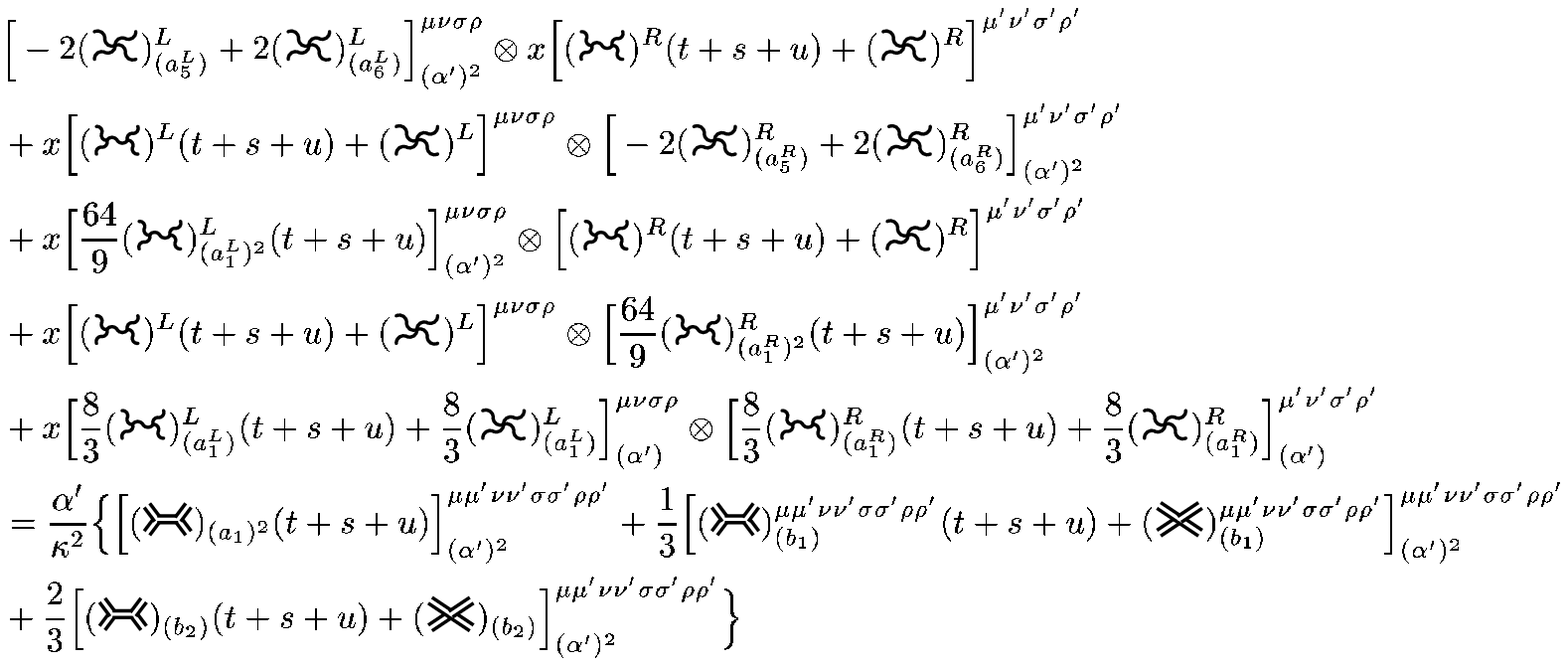}
\caption{The essential diagrammatic relationship between gauge and gravity diagrams.}
\end{figure}

\noindent furthermore we have a relationship generated by the $ca_1$ parts:

\begin{figure}[h]
\includegraphics{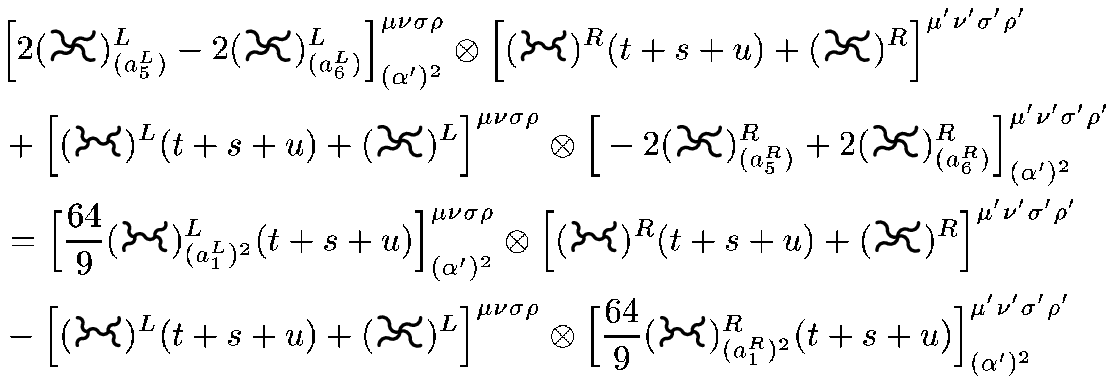}
\caption{Another diagrammatic relationship on the Yang-Mills side between order $\alpha'^2$ and $\alpha'$ operators.}
\end{figure}

These diagrammatical relationships can be readily checked by explicit calculation. It is quite remarkable that the KLT relations provide such detailed statements about pure
Yang-Mills effective field theories, without any reference to gravity at all. To explain the notation in the above diagrammatic statements. An uppercase $L$ or $R$ states that the
scattering amplitude originates from a left or a right mover respectively. The lower indices e.g. $(\alpha')$ and the coupling constants e.g. $(a_3^L)$ or $(a_1^L)^2$ denote respectively
the order of $\alpha'$ in the particular amplitude and the coupling constant prefactor obtained when this particular part of the amplitude is generated from the generic Lagrangian. The parentheses
with $(t+s+u)$ denote that we are supposed to sum over all $t$, $s$ and $u$ channels for this particular amplitude.

\section{Discussion}
It has been observed that it is possible to generalize the KLT open/closed string relations in the effective field theory framework. The KLT relations are seen to
serve as mapping relations between the {\em tree} effective field theories for Yang-Mills and gravity. The belief that general relativity is in fact an effective field theory
at loop orders, makes investigations of its tree level manifestations and connections to other effective actions interesting. Links such as KLT, which are applicable and very simplifying
in actual calculations should be exploited and investigated at the effective Lagrangian level. An important result of our investigations is the generalization of the KLT relations.
It is found that one cannot completely replace the sine function in the KLT relations by an arbitrary function. To order $O(\alpha'^3)$ it is possible
to replace the sine with an odd third order polynomial in $x$. However, the degree of freedom represented by $P_2$ can be completely absorbed into a rescaling of $\alpha'$ at this order, so
additional investigations are required before any conclusions can be drawn.
The mapping relations between the effective theories are found to be broader than those given completely from the KLT relations as the coefficient $P_2$ in the generalized framework
can be chosen freely. Such a rescaling of $P_2$ represents an additional freedom in the mapping. It has been shown that
despite this generalization, the effective extension of the KLT relations is still rather restrictive.

The possibility of an antisymmetric coupling of gravitons needed
in the effective action of a heterotic string has also been
allowed for and is seen to be consistent with the original and
generalized KLT relations. We have learned that detailed
diagrammatic statements can be deduced from the KLT relations.
This presents very interesting aspects which perhaps can be used
to gain additional insight in issues concerning effective
Lagrangian operators. Furthermore we expect this process to
continue, -- at order $O(\alpha'^4)$ we assume that the KLT
relations will tell us about new profound diagrammatic
relationships between effective field theory on-shell operators in
gauge theory and gravity.

The generalized mapping relations represent an effective field theory version of the well known KLT relations. We have used what
we knew already from string theory about the KLT relations to produce a more general description of mapping relations between gravity
and Yang-Mills theory. String theory is not really needed in the effective field theory setting. All that is used here is the tree scattering amplitudes. Exploring if
or if not a mapping from the general relativity side to the gauge theory side is possible produces the generalized KLT relations. The KLT relations could also be considered in
the case of external matter. In this
case, operators as the Ricci tensor and the scalar curvature will not vanish on-shell. Such an approach could perhaps explain more about the KLT relations and introduce additional
aspects in the mapping of operators.

KLT relations involving loops are not yet resolved. One can preform some calculations by making cuts of the diagrams using the
unitarity of the S-matrix but no direct factorizations of loop scattering amplitudes have been seen to support a true loop extension of the KLT relations.
Progress in this direction still waits to be seen, and perhaps such loop extensions are not possible. Loops in string theory and in field theory are not directly comparable,
and complicated issues with additional string theory modes in the loops seem to be unavoidable in extensions of the KLT relations beyond tree level.
Perhaps since 4-point scattering
amplitudes are much less complicated
compared to 5-point amplitudes, it would
be actually more important first to check the mapping
solutions at the 5-point level before considering
the issue of loop amplitudes.

\begin{acknowledgments}
I would like to thank Z. Bern and P.H. Damgaard for useful discussions.
\end{acknowledgments}

\end{document}